\documentclass[a4paper]{jpconf}
\usepackage{graphicx}
\newcommand{\STEREO}{\textsc{Stereo} }
\begin{document}
\title{Search for eV Sterile Neutrinos -- The STEREO Experiment [TAUP 2017]}

\author{Stefan Schoppmann${}^{\ast}$}

\address{Max-Planck-Institut f\"{u}r Kernphysik, Saupfercheckweg 1, 69117 Heidelberg, Germany}
\address{${}^{\ast}$for the \STEREO collaboration}

\ead{stefan.schoppmann@mpi-hd.mpg.de}
\begin{abstract}
In the recent years, major milestones in neutrino physics were accomplished at nuclear reactors: the smallest neutrino mixing angle $\theta_{13}$ was determined with high precision and the emitted antineutrino spectrum was measured at unprecedented resolution. However, two anomalies, the first one related to the absolute flux and the second one to the spectral shape, have yet to be solved. The flux anomaly is known as the Reactor Antineutrino Anomaly and could be caused by the existence of a light sterile neutrino eigenstate participating in the neutrino oscillation phenomenon. Introducing a sterile state implies the presence of a fourth mass eigenstate, while global fits favour oscillation parameters around $\sin^{2}(2\theta)=0.09$ and $\Delta m^{2}=1.8\textrm{eV}^{2}$.

The \STEREO experiment was built to finally solve this puzzle. It is one of the first running experiments built to search for eV sterile neutrinos and takes data since end of 2016 at ILL Grenoble, France. At a short baseline of 10 metres, it measures the antineutrino flux and spectrum emitted by a compact research reactor. The segmentation of the detector in six target cells allows for independent measurements of the neutrino spectrum at multiple baselines. An active-sterile flavour oscillation could be unambiguously detected, as it distorts the spectral shape of each cell's measurement differently.

This contribution gives an overview on the \STEREO experiment, along with details on the detector design, detection principle and the current status of data analysis.
\end{abstract}

\section{Introduction}
The \STEREO experiment is located at the research reactor of the Institut Laue-Langevin (ILL) in Grenoble, France.
It is a reactor antineutrino experiment searching for a light sterile neutrino eigenstate in the eV mass range.
It is thereby searching for a possible explanation of a measured deficit in reactor neutrino flux observed at baselines under 100 metres which is known as Reactor Antineutrino Anomaly (RAA)~\cite{flux}.
Currently, global fits of various experiments hint for a sterile neutrino with oscillation parameters around $\sin^{2}(2\theta)=0.09$ and $\Delta m^{2}=1.8\textrm{eV}^{2}$ as explanation to the RAA~\cite{kopp}.
\STEREO probes this parameter space at a baseline of 9 to 11 metres.

\section{Detector Setup and Experimental Signature}
\begin{figure}[b]
\begin{minipage}{\linewidth}
\includegraphics[width=\linewidth]{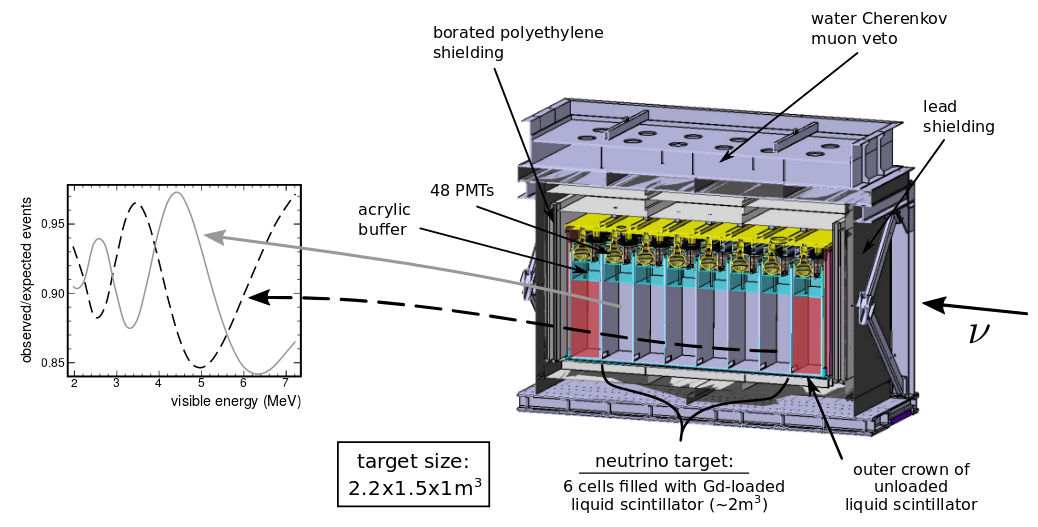}
\caption{\label{fig:detector}Schematic drawing of the \STEREO detector and the principle of sterile neutrino search.}
\end{minipage} 
\end{figure}

Neutrinos are generated in a compact reactor core of 40 cm diameter and 80 cm height and then measured in a liquid scintillator (LS) detector (Fig. \ref{fig:detector}).
The reactor is highly enriched in ${}^{235}$U and runs at a thermal power of 58.3 MW${}_{\textrm{th}}$.
The detection reaction of electron antineutrinos inside the liquid scintillator (LS) of the detector is given by an inverse beta-decay (IBD), which produces a coincidence signal of a positron event followed by a neutron capture event.
48 Hamamatsu 8 inch PMTs observe light produced when the positron annihilation and the radiative neutron capture takes place in the 2000 litre target scintillator as well as the scintillator filled outer-crown volume. 
The kinetic energy of the positron is directly connected to the antineutrino energy, thus allowing for a measurement of the neutrino energy spectrum.
The LS in all volumes are admixtures of LAB, PXE and DIN.
In addition, the target LS is doped with a Gd-$\beta$-diketonate complex to achieve a high neutron detection efficiency.
DIN is added to enhance the particle identification by pulse-shape discrimination.
The spatial resolution of the detector is augmented by optically separating the target volume along the neutrino path of flight in six equally sized cells of 36 cm thickness.
This allows a relative measurement of the neutrino spectrum at six baselines.
\begin{figure}[t]
\includegraphics[width=20pc]{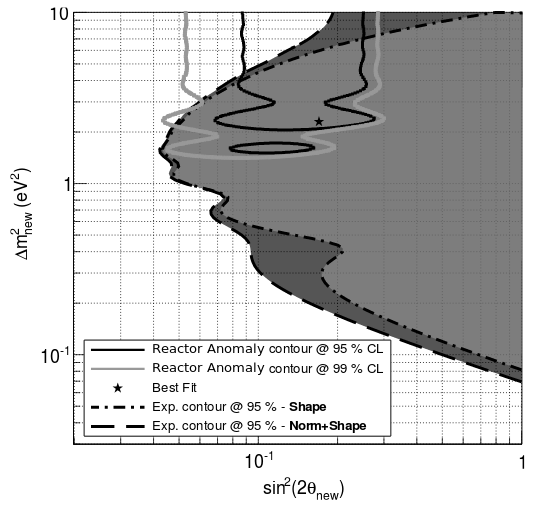}\hspace{2pc}%
\begin{minipage}[b]{14pc}\caption{\label{fig:sensitiv}Projected sensitivity of the \STEREO experiment for one year reactor-on data. The sensitivity study includes the systematic uncertainties from the neutrino reference
spectra, a signal-to-background ratio of 1.5 and detector response contributions and uncertainties
coming from the energy resolution, energy scale uncertainty and detection efficiency.}
\end{minipage}
\end{figure}
The expected sensitivity regions after one year of reactor-on data can be found in Fig. \ref{fig:sensitiv}.

Since the detector is located inside the reactor building, a high flux of neutrons and gammas exist.
Heavy shielding of about 90 tons was installed to suppress those backgrounds.
Shielding materials consist of lead, B${}_{4}$C, borated polyethylene and soft iron as well as a mu-metal layer to shield the detector from external magnetic fields.
In addition, a 15 m.w.e. top-shielding protects against atmospheric radiation.
Remaining atmospheric radiation, neutrons and gammas can mimic the neutrino signature in form of accidental or correlated signal coincidences.
Their magnitude can effectively be estimated by reactor-off phases.
During these periods, a measurement of false neutrino candidates originating only from backgrounds is possible.
Moreover, active countermeasures against background are employed such as the use of a water Cherenkov muon veto above the neutrino detector, a specific set of cuts on the PMT hit pattern of each event and selections on the topology of the coincidence signal or the pulse-shape discrimination (PSD) parameter.

\section{Calibration and Detector Performance}
Calibration of the PMT and electronics response is done using an LED light injection system in the target and the outer-crown.
Moreover, three calibration systems for the deployment of radioactive sources exist:
an internal tube system is installed in several target cells, an external system allows to position sources in any height around the detector, but inside the shielding and a tube below the detector allows calibration from underneath.

The detection efficiency is determined by an AmBe gamma-neutron source.
The capture time of neutrons is found to be stable over time at (16.2 $\pm$ 0.2) $\mu s$ and it is in agreement with the capture time from neutrino candidate events of (16.5 $\pm$ 0.6) $\mu s$.
From the AmBe calibration data, it is also possible to infer an excellent energy containment by observing the capture peaks of neutrons by hydrogen and gadolinium with good resolution.
The fraction of neutron captures by gadolinium is determined as 86\% at the centre of the target.

The energy scale calibration and non-linearity estimation is achieved by the utilisation of gamma sources between 0.5 and 4.4 MeV.
For this analysis, cell-to-cell variations in acceptance and light-crosstalk between neighbouring cells are taken into account. Thus, the measured charge $Q_{i}$ is assumed as the sum of energy depositions in all cells:
\begin{equation}
Q_{i} = \alpha_{i} \sum_{j=cells} E^{dep}_{j} \cdot f_{j} \cdot L_{j \rightarrow i} \textrm{.}
\end{equation}
In this equation, $E^{dep}_{j}$ denotes the energy deposited in cell $j$, $f_{j}$ denotes the light yield in cell $j$, $L_{j \rightarrow i}$ the light-crosstalk from cell $j$ to $i$ and $\alpha_{i}$ the light-acceptance of cell $i$.
All appearing coefficients are obtained in-situ from calibration data as:
\begin{equation}
Q_{i} = \sum_{j=cells} E^{dep}_{j} \cdot C_{j} \cdot L_{j \rightarrow i}
\end{equation}
where $C_{j}$ denotes the measurable combination of the above parameters.

\begin{figure}[t]
\begin{minipage}{\linewidth}
\includegraphics[width=\linewidth]{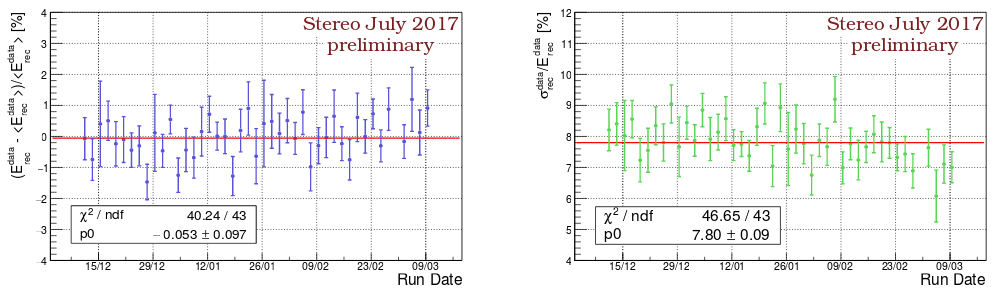}
\caption{\label{fig:hydrogen}Position (left) and width (right) of energy depositions from neutron captures by hydrogen over time.}
\end{minipage} 
\end{figure}

The light yield is determined as 270 PE per MeV and a good homogeneity of the energy response is found across the fiducial volume using a ${}^{54}$Mn gamma source.
A small 3\% difference is found when comparing the energy response at the border with that of the centre.
The findings are reproduced with simulations at the percent level.
Furthermore, the response is found stable over time and a good resolution is found by investigating captures of spallation neutrons by hydrogen throughout the entire target volume (Fig. \ref{fig:hydrogen}).

\begin{figure}[b]
\includegraphics[width=24pc]{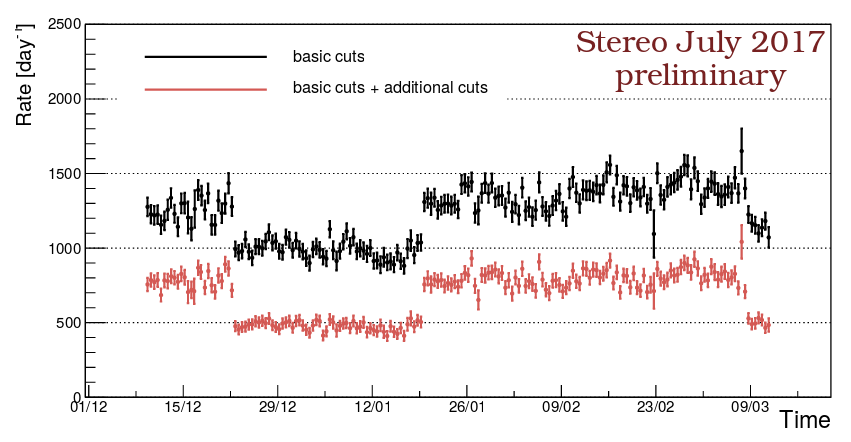}\hspace{2pc}%
\begin{minipage}[b]{10pc}\caption{\label{fig:rates}Neutrino candidate rates for different event selections as explained in the text.}
\end{minipage}
\end{figure}

The neutrino candidate event rates over time can be found in Fig. \ref{fig:rates}.
All rates are corrected for dead times and their dependence on atmospheric pressure.
The change in rate for reactor-on and reactor-off periods is clearly seen.
The black data points are obtained by requiring for the total energy of the prompt event 1.5 MeV $< E <$ 8 MeV and less than 1.1 MeV deposited in the outer-crown.
The delayed event is required to have an energy between 5 and 10 MeV in the entire detector and more than 1 MeV has to be detected in the target.
Moreover, the delay between the prompt and delayed events may not exceed 70 μs.
For the prompt event signal, the PSD parameter is required to deviate less than 2.5 $\sigma$ from the central PSD value for that event energy.

The red data points depict the rates when using an additional set of more stringent cuts.
These are defined by taking into account the event topology of the PMT hit pattern in order to reduce the amount of background coincidences caused by stopping muons.
Moreover, accidental coincidences are suppressed by a large fraction when requiring a distance between the prompt and the delayed events of less than 0.4 m.
A further reduction of non-IBD events can be achieved by requiring a highly localised energy deposition. This is done by demanding less than 0.7 MeV of energy in the cells next to the main cell of the interaction.
Applying the entire set of cuts yields a factor of two improvement on the signal-to-background ratio while introducing a signal inefficiency in the few percent order.

When comparing reactor-on and reactor-off data, it is found that the shielding is sufficient: accidental background rates are below the design level and no increase of correlated events by neutrons from the reactor is found.
Furthermore, a neutrino candidate rate of 300 neutrinos per day is found, well in agreement with expectations.

\section{Summary and Outlook}
Since its start of data taking in November 2016, the \STEREO experiment has collected 70 days of reactor-on and 25 days of reactor-off data.
The ongoing analysis of this data testifies a good detector performance and is currently further refined, e.g. by cut optimisation and more detailed cosmic background studies.
Following the finalisation of the energy scale, an oscillation analysis is going to be performed.
At the same time, another 150 days of reactor-on data taking is upcoming until 2018, allowing for improved statistics.


\section*{References}
\begin{thebibliography}{9}
\bibitem{flux} Mention G, Fechner M, Lassere T, Mueller T A, Lhuillier D, Cribier M and Letourneau A 2011 {\it Phys. Rev. D} {\bf 83} 073006 
\bibitem{kopp} Kopp J, Machado P A N, Maltoni M and Schwetz T 2013 {\it JHEP} {\bf 1305} 050 
\end{thebibliography}
\end{document}